\documentclass[aps,prl,reprint,groupedaddress]{revtex4-1}

\usepackage[colorlinks,citecolor=red,urlcolor=blue]{hyperref}
\usepackage{graphicx}
\graphicspath{{figures/final/}}

\begin{document}


\title{\textit{Real-time} X-ray Monitoring of the Nucleation and Growth of AlN Epitaxial Films on Sapphire (0001)}


\author{Guangxu Ju}
\email[Author to whom correspondence should be addressed. Electronic mail:]{gju@anl.gov}
\affiliation{Materials Science Division, Argonne National Laboratory, Argonne, IL 60439}

\author{Matthew J. Highland}
\affiliation{Materials Science Division, Argonne National Laboratory, Argonne, IL 60439}

\author{Jeffrey A. Eastman}
\affiliation{Materials Science Division, Argonne National Laboratory, Argonne, IL 60439}

\author{Rebecca Sichel-Tissot}
\affiliation{Materials Science Division, Argonne National Laboratory, Argonne, IL 60439}

\author{Peter M. Baldo}
\affiliation{Materials Science Division, Argonne National Laboratory, Argonne, IL 60439}

\author{Peter Zapol}
\affiliation{Materials Science Division, Argonne National Laboratory, Argonne, IL 60439}

\author{Hua Zhou}
\affiliation{X-ray Science Division, Argonne National Laboratory, Argonne, IL 60439}

\author{Carol Thompson}
\affiliation{Department of Physics, Northern Illinois University, DeKalb IL 60115}

\author{Paul H. Fuoss}
\email{Current address: SLAC National Accelerator Laboratory, Menlo Park, CA 94025 USA}
\affiliation{Materials Science Division, Argonne National Laboratory, Argonne, IL 60439}


\date{\today}

\begin{abstract}
We report the results of x-ray scattering studies of AlN on $c$-plane sapphire during reactive radiofrequency magnetron sputtering. The sensitivity of \textit{in situ} x-ray measurements allowed us to follow the structural evolution of strain and roughness from initial nucleation layers to fully relaxed AlN films. A growth rate transient was observed, consistent with the initial formation of non-coalesced islands with significant oxygen incorporation from the substrate. Following island coalescence, a steady state growth rate was seen with a continuous shift of the $c$ and $a$ lattice parameters towards the relaxed bulk values as growth progressed, with films reaching a fully relaxed state at thicknesses of about 30 nm.
\end{abstract}


\maketitle

Aluminum nitride (AlN) is often used as the initial growth layer for wide-bandgap semiconductor device structures, such as deep-ultraviolet light emitting diodes (LEDs) and heterostructure field-effect transistors.\cite{akasaki2006breakthroughs,amano2016development}  Typically, AlN is grown on poorly lattice matched sapphire (corundum-structured Al$_{2}$O$_{3}$) substrates, resulting in a dense and complex network of defects.\cite{dovidenko1996aluminum,kehagias2001misfit,tokumoto2008high} These defects propagate into the device structures act as non-radiative recombination centers where they decrease the internal quantum efficiency in the active region of LEDs.\cite{schubert2007effect} Thus, understanding the formation pathways of the initial AlN nucleation layer is important in optimizing the growth of the heterostructures with low defect density and improved device performance.

Structures formed in the early stages of growth play an important role in controlling the orientation, strain state, and defect content of heteroepitaxial AlN films, and thus have been the subject of several recent studies.\cite{kang2001x, wang2006real,headrick1998ion,imura2006high,moustakas2013role,li2015growth} For example, previous experimental investigations of the growth mechanism revealed that a planar two-dimensional AlN epilayer can be grown on large lattice mismatch (-13.2$\%$ strain) sapphire (0001) substrates, possibly due to an extended atomic distance mismatch (EADM) configuration.\cite{sun1994crystallographic,kung1994crystallography,banal2008initial,shim1998evolution} The crystallographic orientation of AlN relative to sapphire is (0001)$\parallel$(0001) and[10$\overline{1}$̅0]$\parallel$[11$\overline{2}$0], i.e., the AlN lattice is rotated 30$^{\circ}$ about the [0001] axis with respect to the sapphire lattice.\cite{dovidenko1996aluminum,sun1994crystallographic} Kang et al.\cite{kang2001x} used \textit{ex situ} high-resolution synchrotron x-ray diffraction observations to investigate strain development in AlN films grown on (0001) sapphire. They reported that 8.5 nm thickness planar films were strained in-plane about 2$\%$ in compression when observed at room temperature, and that complete relaxation of films to strain-free states was not observed until films reached thicknesses exceeding 100 nm. They attributed metastability of very thin films in a strained state as being due to a ``ten-to-eleven" domain-matching configuration. They also found that nonuniform strain distributions appeared in thicker films and that the surface morphology transitions from a planar layer to islands near a film thickness of $\sim$25 nm, consistent with their observed surface roughness evolution. Wang and coworkers\cite{wang2006real} used \textit{in situ} synchrotron x-ray diffraction and reflectivity measurements to investigate strain relaxation of the nitridation layer on $c$-plane sapphire in an ultrahigh vacuum chamber using surface x-ray diffraction. They found that heteroepitaxial strain in the surface nitride layer ($\sim$2.5 \text{\AA} thickness) is significantly different than that of both the substrate and bulk AlN, with an in-plane lattice parameter approximately 1.5$\%$ smaller than that of bulk AlN. However, they did not study subsequent AlN growth on this nitride layer. Despite the progress in these previous studies, there are many open questions related to the earliest stages of AlN growth behavior, particularly during the first 100 \text{\AA} of growth.

In the following, we describe a \textit{real-time} synchrotron x-ray scattering study, focusing on characterizing growth rate, roughness, and strain during the first few hundred \text{\AA}ngstroms of epitaxial growth of AlN on (0001) sapphire, correlating those important parameters with film thickness. Using x-ray reflectivity,\cite{headrick1998ion,fuoss1989atomic,ju2014continuous,ju2017role} we measured the film thickness and growth modes, and also determined the evolution of three-dimensional strain using both in-plane grazing incidence x-ray (GIX) scattering and out-of-plane specular crystal truncation rod (CTR) observations. Our results reveal the evolution of the strain and growth rate during heteroepitaxial growth.


Films were grown in a custom-built, on-axis sputtering chamber mounted on a five-circle diffractometer at Sector 12ID-D of the Advanced Photon Source (APS).\cite{folkman2013modular} A one-inch diameter aluminum target (99.999$\%$ purity) was mounted and operated at an RF power of 25 W. The approximately 3.5 L chamber volume was typically maintained at 15 mTorr pressure while flowing 0.3 slpm of a 70$\%$ Ar, 30$\%$ N$_{2}$ (to provide the reactive nitrogen) mixture of ultrahigh purity grade gasses. In order to minimize contaminants while growing AlN, the vacuum chamber was baked, the target was presputtered in an Ar atmosphere, and aluminum was sputter deposited on the chamber walls prior to loading substrates.

Single-crystal 10$\times$10$\times$1 mm$^{3}$ $c$-plane oriented sapphire substrates were used. Each substrate was ultrasonically cleaned with organic solvents (acetone and methanol), rinsed with deionized water, and dried with compressed high purity nitrogen gas.\cite{cuccureddu2010surface,fu2006hydroxylated} The substrates were then annealed at 1100 $^{\circ}$C in pure O$_{2}$ for two hours in a tube furnace. Atomic force microscopy (AFM) on the post-annealed substrates found ordered surface step morphologies with clearly defined terraces, a miscut angle of 0.15$^{\circ}$ -- 0.28$^{\circ}$, and a step height of 0.22 \text{\AA}. Prior to growth, the sapphire substrates were further annealed in the growth chamber in N$_{2}$/Ar gas and Ar$^{+}$/N$^{+}$ sputtering species at 830 $^{\circ}$C for approximately four hours to eliminate OH species from the substrate surface to obtain an Al termination layer,\cite{fu2006hydroxylated,eng2000structure} and to ensure reproducible results.  After this annealing, the substrates were cooled to 700 $^{\circ}$C  for growth.

In order to access a wide range of reciprocal space with minimum angular motion, an x-ray energy of 28 keV ($\lambda$ = 0.4428 \text{\AA}) was used. X-rays scattered by the sample were detected using a PILATUS-100K-S area detector located one meter from the sample. The detector's 450 $\mu$m thick silicon sensor had a 15$\%$ quantum efficiency for 28 keV photons. 

\begin{figure}[h]
\includegraphics[width=0.7\columnwidth]{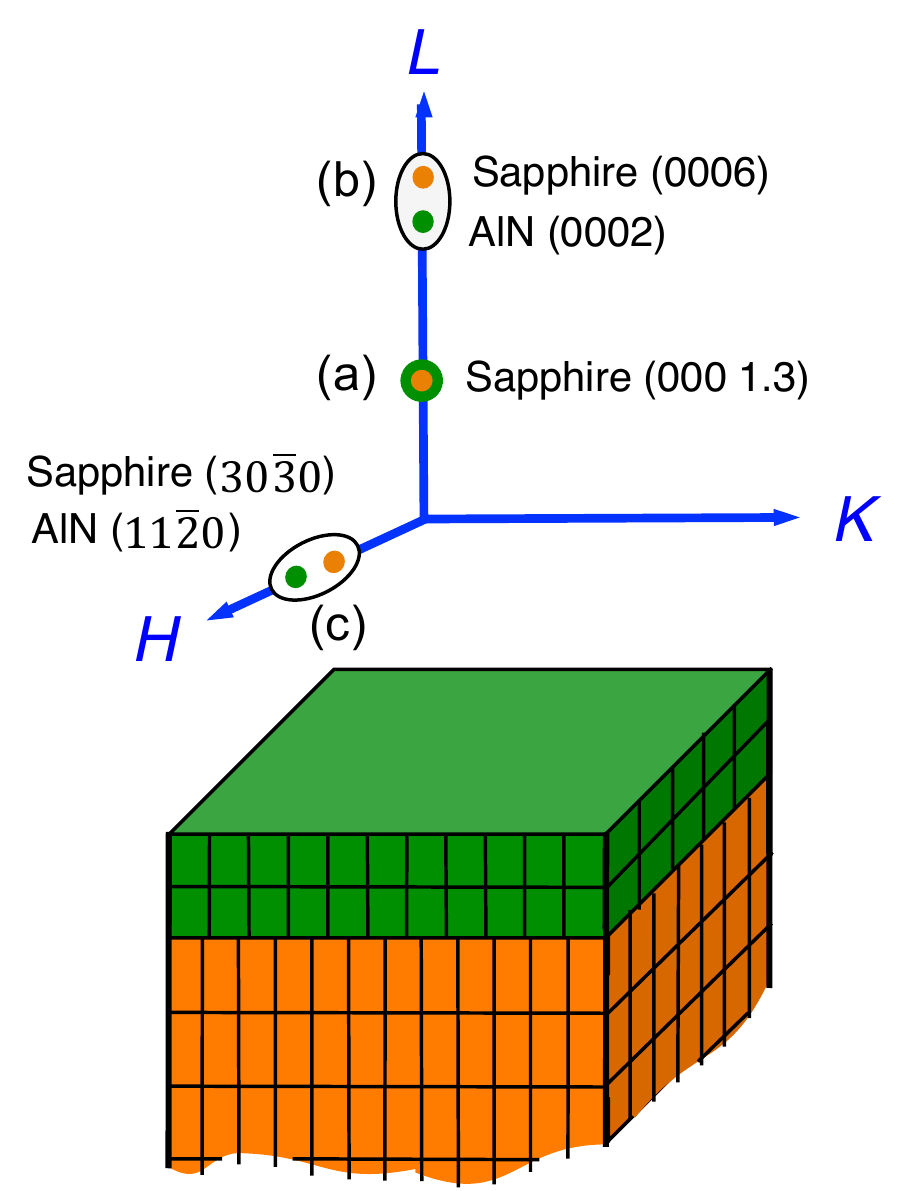}
\caption{\label{Geometry} Schematic showing the three regions of reciprocal space monitored during separate AlN growths: (a) specular CTR at the $L$=1.3 position of sapphire, which is the (000 $\frac{1}{2}$) position of bulk AlN at room temperature; (b) the specular CTR in the region of the AlN (0002) and sapphire (0006) Bragg reflections, and (c) the in-plane region near the sapphire (30$\overline{3}$̅0) and AlN (11$\overline{2}$̅0) Bragg peaks.  } 
\end{figure}

During each growth, one of three regions of reciprocal space was monitored, as shown in Fig. \ref{Geometry}(a). During growth of some AlN films, we continuously monitored the intensity of the (000$L$) specular CTR at the sapphire $L$=1.3 position, which is approximately the (000 $\frac{1}{2}$) position of bulk AlN at room temperature. For simplicity, we refer to the room-temperature (RT) sapphire substrate reciprocal lattice\cite{kung1994crystallography} ($c_{\text{sapphire}}^{\text{RT}}$=1.2991 nm and $a_{\text{sapphire}}^{\text{RT}}$=0.4748 nm) in indexing diffraction positions, regardless of the sample temperature. The lattice parameters of bulk AlN are $c_{\text{AlN}}^{700}$  =0.4992 nm and $a_{\text{AlN}}^{700}$ =0.3121 nm at 700 $^{\circ}$C, and $c_{\text{AlN}}^{\text{RT}}$ =0.4981 nm and $a_{\text{AlN}}^{\text{RT}}$  =0.3112 nm at RT.\cite{reeber2001high,yim1974thermal} For other samples, during growth we instead monitored either the in-plane scattering around the AlN (11$\overline{2}$̅0) Bragg peak or the specular CTR in the region of the AlN (0002) Bragg reflection. Data were fit to a pseudo-Voigt function, including a linear background, to identify each peak's centroid position and full width at half maximum (FWHM). The $a$ lattice parameter at the final film thickness was determined from the difference between the centroids of the AlN (11$\overline{2}$̅0) and (22$\overline{4}$̅0) Bragg peaks. 

\begin{figure}[h]
\includegraphics[width=\columnwidth]{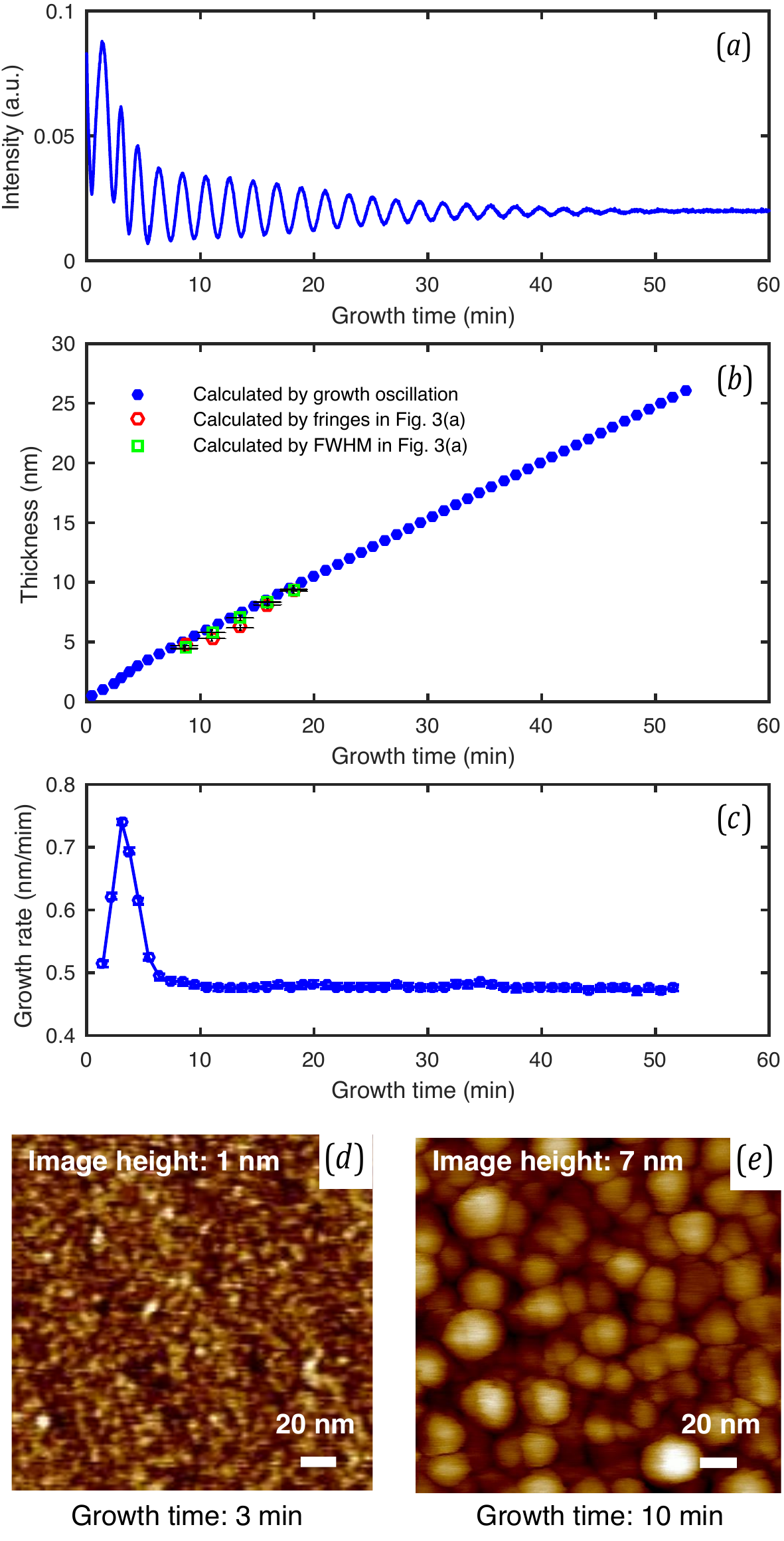}
\caption{\label{Growth_oscillations} (a) \textit{Real-time} monitoring the specular reflectivity on sapphire (000$L$) at $L$=1.3 during AlN deposition at 700 $^{\circ}$C. (b) The interpreted thickness as function of the growth time. (c) The oscillation rate with growth time, which is derived from the growth time divided by the film thickness. (d) 200$\times$200 nm$^2$ $\textit{ex situ}$ AFM images of 3 min- and (e) 10 min-AlN growth. } 
\end{figure}

\begin{figure}[h]
\includegraphics[width=\columnwidth]{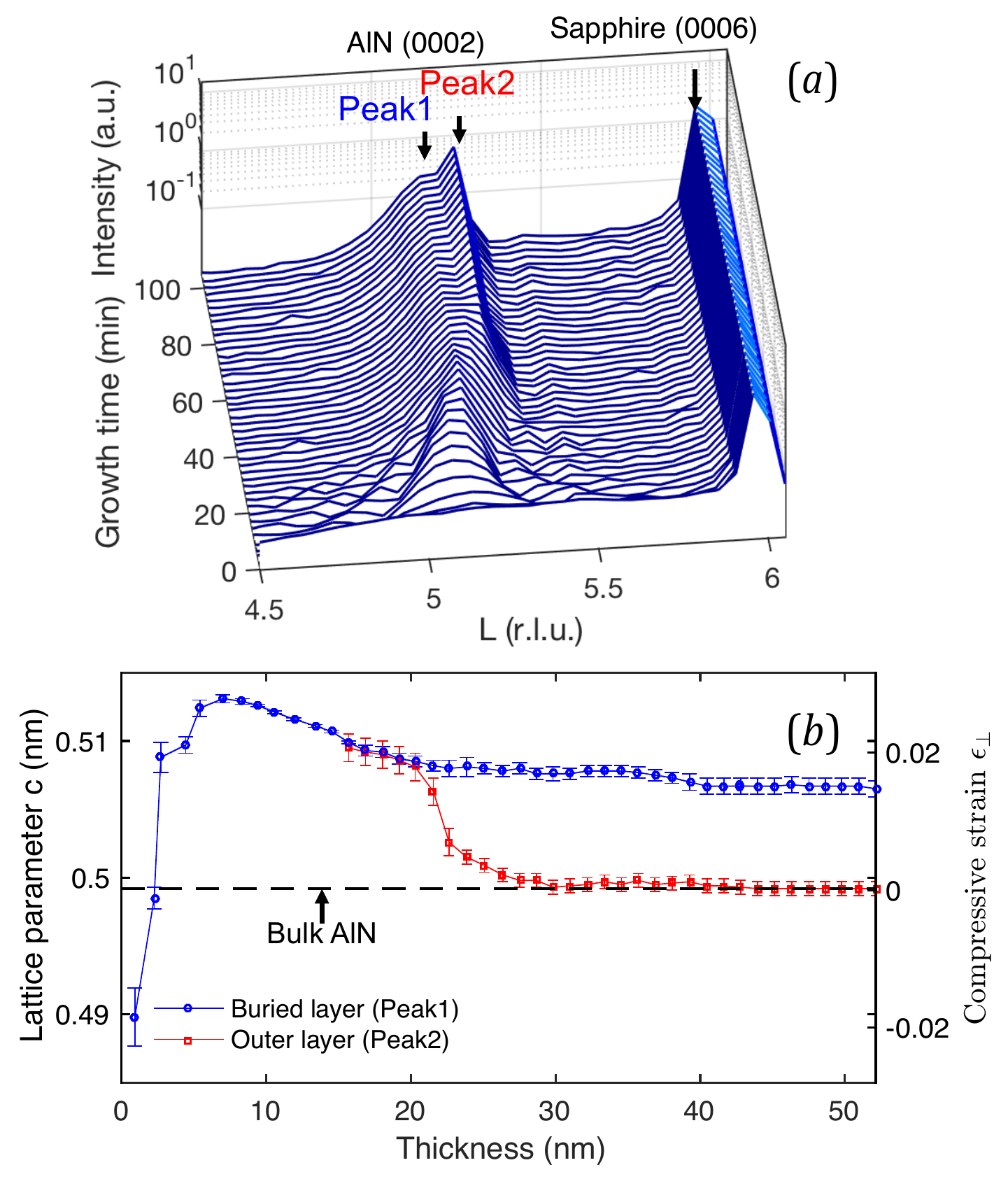}
\caption{\label{Out-of-plane} (a) The time-evolution of repeated 000$L$ specular rod scans near the AlN (0002) Bragg peak during growth. Each scan took approximately 145 seconds. (b) The corresponding $c$ lattice parameter and compressive strain evolution with film thickness extracted from (a) using growth rate data shown in Fig. \ref{Growth_oscillations}(b). The dashed line is the literature value for the $c$ lattice parameter of bulk AlN at 700 $^{\circ}$C.} 
\end{figure}

Figure \ref{Growth_oscillations}(a) shows the $L$=1.3 CTR intensity as a function of time. Heteroepitaxial intensity oscillations were observed with a period corresponding to 1 nm per oscillation.\cite{ju2014continuous} The thickness change per oscillation was determined by measuring the period of the Kiessig fringes in the vicinity of the (0002) AlN Bragg reflection, seen in Fig. \ref{Out-of-plane}(a). The loss of resolvable intensity oscillations at a growth time of $\sim$54 min is evidence of surface roughening. 

Figure \ref{Growth_oscillations}(b) shows the evolution of the film thickness with growth time. As seen in Fig. \ref{Growth_oscillations}(c), a transient growth rate was observed during the first few unit cells of growth and then a steady state growth rate of 0.48 nm/min was observed after films reached a thickness of $\sim$7 nm (at $\sim$14 min).  Film surface morphologies were observed by AFM, as shown in Fig. \ref{Growth_oscillations}(d, e) after growth times of three and ten minutes, respectively. These images show that a $\sim$5nm thick film was not fully coalesced.\cite{imura2006high,hashimoto1998structural,sun2013situ} Thus, the apparently faster growth rate observed during the first several minutes of growth may have occurred because films were not yet fully coalesced (for example, if material is selectively increasing island height rather than uniformly spreading on surface).  After film coalescence was completed, a constant growth rate was observed. The transient growth rate behavior observed is similar to observations by Headrick and coworkers\cite{headrick1998ion} of significantly faster-than-steady state growth during initial ion-assisted GaN growth.

In order to study the evolution of three-dimensional strain in the growing AlN films, we monitored both out-of-plane and in-plane regions of reciprocal space near film Bragg peaks. Figure \ref{Out-of-plane}(a) shows the time evolution of the specular AlN (0002) Bragg peaks and Fig. \ref{Out-of-plane}(b) shows the out-of-plane $c$ lattice parameter versus film thickness extracted from Fig. \ref{Out-of-plane}(a) using growth rate data shown in Fig. \ref{Growth_oscillations}(b). Upon island nucleation, we observed an $\sim$2$\%$ smaller film $c$ lattice parameter than that of bulk AlN, but the (0002) plane spacing increased rapidly as growth proceeded, reaching a value more than 2$\%$ larger than that of bulk AlN at a film thickness of 7 nm. As growth continued beyond 7 nm film thickness, the $c$ lattice parameter began to relax towards the bulk AlN value. Beginning at a film thickness of $\sim$20 nm, two AlN (0002) Bragg peaks could be resolved, indicating that two distinct strain states (two $c$ lattice parameters) had developed. Figure \ref{Out-of-plane}(b) shows the resulting strain evolution derived from both peaks. As the film thickness continued to increase to 50 nm, little further change was observed in the larger of the two $c$ lattice parameters (obtained from Peak 1), but the second lattice parameter (obtained from Peak 2) decreased by approximately 1.5$\%$ as the film thickness increased from 20 to 30 nm and then remained constant during further growth. Our interpretation is that Peak 1 results from a thin buried layer that maintains a larger $c$ lattice parameter than bulk AlN, while Peak 2 arises from the outer portion of the film, which has a $c$ lattice parameter that relaxes, reaching the fully relaxed bulk value at a film thickness of about 30 nm. The larger observed width of Peak 1 compared with Peak 2 is also consistent Peak 1 arising from a thin region of the sample near the film / substrate heterointerface.

\begin{figure}[h]
\includegraphics[width=\columnwidth]{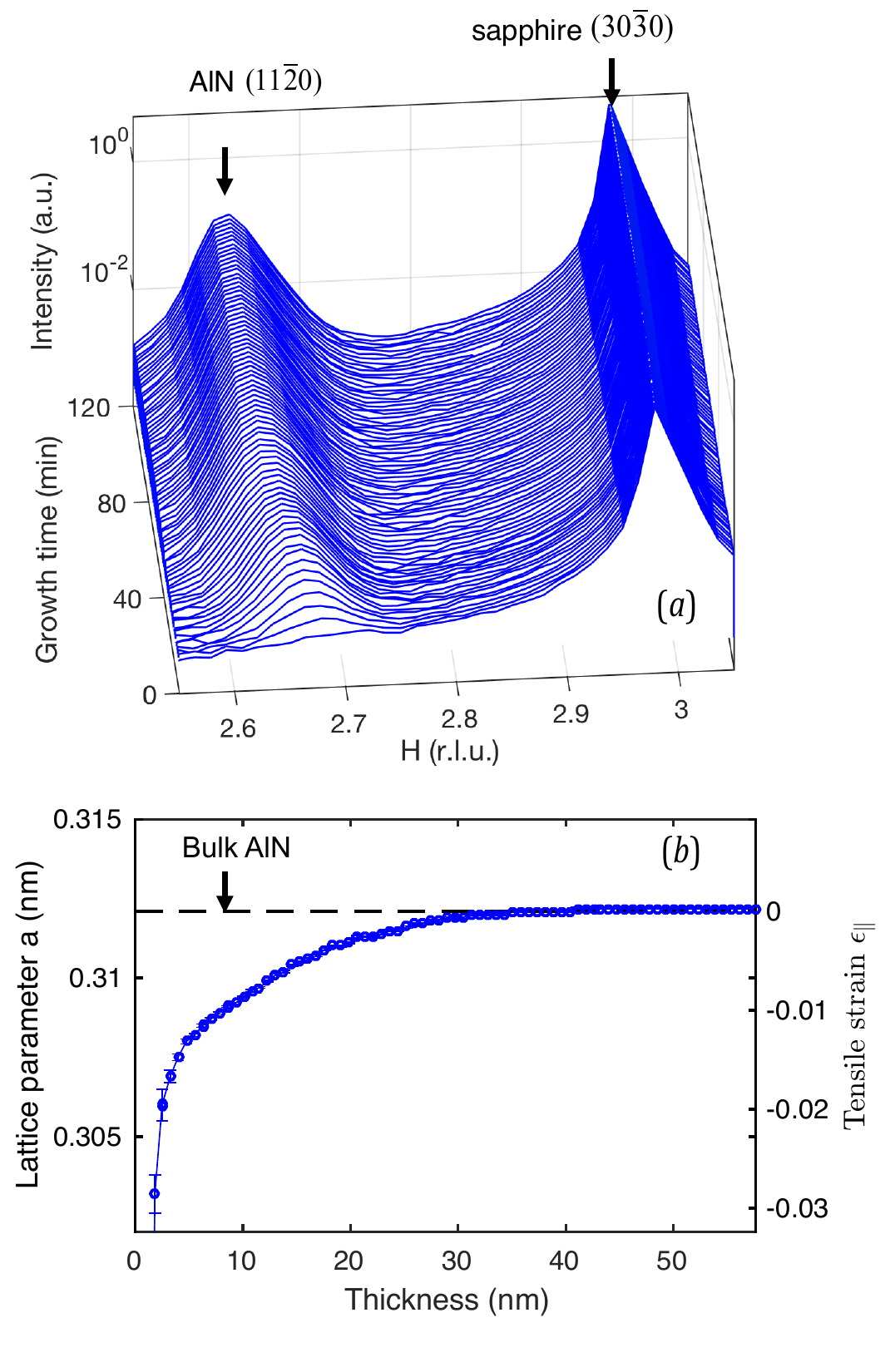}
\caption{\label{In-plane} (a) The time-evolution of repeated in-plane profiles through the AlN (11$\overline{2}$̅0) peak during growth. Each scan took approximately 95 seconds. (b) $a$ lattice parameter and tensile strain with film thickness extracted from (a) using growth rate data shown in Fig. 2(b). The dashed line is the literature value for $a$ lattice parameter of bulk AlN at 700 $^{\circ}$C.  } 
\end{figure}

Figure \ref{In-plane}(a) shows the time evolution of the in-plane AlN (11$\overline{2}$̅0) Bragg peak and Fig. \ref{In-plane}(b) shows the in-plane $a$ lattice parameter versus film thickness extracted from Fig. \ref{In-plane}(a) using growth rate data shown in Fig. \ref{Growth_oscillations}(b). Films were initially observed to have a smaller in-plane lattice constant than bulk AlN, but the value fully relaxed to the bulk in-plane lattice parameter as the film approached 30 nm thickness. 

Both the $a$ and $c$ lattice parameters were observed to be smaller than bulk AlN values for film thicknesses less than 2.5 nm. This is inconsistent with the behavior expected if initial AlN non-coalesced islands were coherently strained to the substrate, which would result in a smaller-than-bulk in-plane lattice constant and a correspondingly larger-than-bulk out-of-plane lattice constant due to the Poisson effect. Our hypothesis is that the initial smaller-than-bulk values of \textit{both} lattice constants is due to initial incorporation of oxygen from the substrate into the nucleation layer, consistent with numerous reports that oxygen incorporation into AlN decreases the lattice volume.\cite{hashimoto1998structural,mohn2016polarity,sun2017influence,slack1973nonmetallic}

Because the substrate is the oxygen source, oxygen incorporation into the film is likely to decrease rapidly after the first few monolayers of growth, particularly after the films become fully coalesced. Our data indicate that as growth progresses beyond a thickness of 2.5 nm, the (0002) d-spacing rapidly increases to a value larger than that of the bulk AlN value, while the in-plane lattice constant remains smaller than the bulk AlN value. This is consistent with the behavior that would be expected if the growing coalesced films were initially coherently strained compressively to the oxygen-containing AlN islands. As growth continues, the films then relax towards having bulk $a$ and $c$ lattice parameters, reaching the fully relaxed state at film thicknesses of about 30 nm. Compared with previous \textit{ex situ} studies of films at room temperature,\cite{kang2001x,shim1998evolution} we observed relaxation to zero strain state at much smaller film thicknesses and saw no indication of kinetic stabilization of strain due to the formation of domain matching configurations. This difference in behavior is likely due to our observations of the films at the 700 $^{\circ}$C  growth temperature, where relaxation processes were not kinetically limited.

In summary, we found that during AlN nucleation, both $a$ and $c$ lattice parameters were smaller than bulk values, possibly indicating some initial oxygen incorporation from the substrate into AlN islands. The observed transient initial growth rate is consistent with initial island nucleation behavior. Following island coalescence, a rapid increase in the $c$ lattice constant is observed, consistent with the oxygen-free films being constrained by in-plane matching to the oxygen-containing islands. With further growth, both $a$ and $c$ lattice parameters continuously evolved towards bulk values as growth proceeded, reaching fully relaxed bulk AlN values at film thicknesses of approximately 30 nm. The results indicate the importance of possible initial incorporation of oxygen from the substrate in influencing subsequent growth behavior.

This research was supported by the U.S. Department of Energy (DOE), Office of Science, Office of Basic Energy Sciences, Division of Materials Science and Engineering. The use of the Advanced Photon Source is supported by the U.S. Department of Energy, BES under Contract No. W-31-109-ENG-38.

\bibliography{AlN_final}

\end{document}